\documentclass[aps,prl,twocolumn,groupedaddress,amsmath,amssymb,showpacs]{revtex4}
\usepackage{graphicx}
\usepackage{amsmath}
\usepackage{dcolumn}% Align table columns on decimal point
\usepackage{bm}% bold math
\newcommand{\bx}{\mathbf{x}}

\newcommand{\bz}{\mathbf{z}}

\newcommand{\bq}{\mathbf{q}}
\newcommand{\bk}{\mathbf{k}}
\newcommand{\br}{\mathbf{r}}

\newcommand{\enve}{\hat{\bm \Psi}}
\newcommand{\bA}{\mathbf{A}}

\newcommand{\im}{\mathrm{i}}

\newcommand{\abA}{\hat{\mathbf{A}}}

\newcommand{\ha}{\hat{ a} }

\newcommand{\ra}{\rangle}
\newcommand{\la}{\langle}

\newcommand{\vepsilon}{\bm \epsilon}

\newcommand{\ve}{\bm e}

\begin{document}
\title{Exact Paraxial Quantization}
\author{A. Aiello}
\author{J.P. Woerdman}
\affiliation{Huygens Laboratory, Leiden University\\
P.O.\ Box 9504, 2300 RA Leiden, The Netherlands}
\begin{abstract}
A non-perturbative quantization of a paraxial electromagnetic
field is achieved via a generalized dispersion relation imposed on
the longitudinal and the transverse components of the photon wave
vector. This new theoretical formalism yields a seamless
transition between the paraxial- and the Maxwell-equation
solutions. This obviates the need to introduce either \emph{ad
hoc} or perturbatively-defined field operators. Moreover, our
(exact) formalism remains valid beyond the quasi-monochromatic
paraxial limit.
\end{abstract}
 \maketitle
%
%03.65.Db    Functional analytical methods in quantum mechanics
%
%03.65.Ca    Formalism in quantum mechanics
%
%03.70.+k    Theory of quantized fields
%
%41.20.Jb    Electromagnetic wave propagation; radiowave propagation
%
%41.85.-p    Beam optics
%
%03.65.Ud    Entanglement and quantum nonlocality (e.g. EPR
%paradox, Bell's inequalities, GHZ states, etc.) (for entanglement
%production in quantum information, see 03.67.Mn; for entanglement
%in Bose-Einstein condensates, see 03.75.Gg)
%
%03.65.Nk    Scattering theory
%
%03.67.Mn    Entanglement production, characterization, and
%manipulation (see also 03.65.Ud Entanglement and quantum
%nonlocality; for entanglement in Bose-Einstein condensates, see
%03.75.Gg)
%
%42.25.Ja    Polarization
%
%42.50.Dv    Nonclassical states of the electromagnetic field,
%including entangled photon states; quantum state engineering and
%measurements (see also 03.65.Ud Entanglement and quantum
%nonlocality, e.g. EPR paradox, Bell's inequalities, GHZ states,
%etc.)
%
%
In this Letter we seek an answer to the question: What is the
quantum-mechanical state of a photon in a beam of light whose
propagation can be classically described by a paraxial wave
equation?
The problem of quantum propagation of paraxial fields was
considered by several authors in the past \cite{Paraxia}. In these
contributions the approach was either \emph{ad hoc} or
approximated. A notable exception was the work of Deutsch and
Garrison where the authors developed a perturbative quantization
scheme \cite{Deutsch91a}. However, although their theory  formally
solves, order by order in a perturbation expansion, the problem of
paraxial quantum propagation, it suffers from two main
limitations. First, it does not provide any clear and easily
manageable formula for the paraxial quantum modes. Second, it
requires the quasi-monochromatic approximation which is unsuitable
for, e.g., the description of  the very quantum phenomenon of
propagation of broad-band entangled photons provided by
spontaneous parametric down conversion (SPDC).
The increasing importance, for quantum information in general
\cite{NielsenBook} and quantum cryptography \cite{Gisin02} in
particular, of a proper description of paraxial propagation of
SPDC entangled photons, calls for an exact (namely,
non-perturbative) theory of paraxial quantum fields.
%
%We believe that this situation is at best
%unsatisfactory, at worst highly dangerous.
%

The aim of this Letter is to introduce a non-perturbative method
to quantize a paraxial electromagnetic field. Our scheme is
conceptually simple: we begin by considering the quantized
transverse electromagnetic vector potential calculated at the
(arbitrary) initial time $t=0$. Then we select from the
 wave-vector space  only those field
configurations which are \emph{exact} solutions of the paraxial
wave equation, and let them evolve in time according to the
d'Alembert wave equation. This procedure automatically ensures the
validity of both the canonical commutation relations and the
transversality conditions for the fields. We stress that our
theory accounts for the  \emph{axial} propagation of vector fields
with \emph{any} spectral and spatial bandwidth and reproduces the
well known \emph{paraxial} results in the limit of narrow
beam-like fields.

%Furthermore, our theory applies to fields with arbitrary
%bandwidth.
%

To begin  with, let us consider the vector potential operator
$\hat{\bA}(\br,t)= \hat{\bA}^{(+)}(\br,t) +
\hat{\bA}^{(-)}(\br,t)$ in the Coulomb gauge which  can be written
in the plane-wave basis as \cite{LoudonBook}
\begin{equation}\label{a20}
\begin{array}{ll}
  \displaystyle{ \hat{\bA}^{(+)}(\br,t)  = } & \displaystyle{ \int \mathrm{d}^3 \bk
  \left( \frac{\hbar}{16 \pi^3 \varepsilon_0 c |\bk|} \right)^{1/2}}\\
   & \displaystyle{\times  \sum_{\lambda = 1}^2 \vepsilon^{(\lambda)}(\bk)
  \hat{a}_\lambda(\bk) \exp \left( \im \bk \cdot \br - \im c |\bk| t \right)} ,\\
\end{array}
\end{equation}
and $\abA^{(-)}(\br,t)$ is the Hermitian conjugate of
$\abA^{(+)}(\br,t)$. It clearly satisfies the d'Alembert wave
equation $\square \hat{\bA}(\br,t) =0$. The two unit polarization
vectors $\vepsilon^{(\lambda)}(\bk), (\lambda = 1,2)$ are
transverse $\vepsilon^{(\lambda)}(\bk) \cdot \bk = 0$ and mutually
orthogonal $ \vepsilon^{(1)}(\bk) \cdot \vepsilon^{(2)}(\bk) = 0$.
Moreover, the annihilation and creation operators
 satisfy the canonical
commutation rules $
[\ha_{\lambda}(\bk),\ha_{\lambda'}^\dagger(\bk')] =
\delta_{\lambda \lambda'}\delta^{(3)}(\bk - \bk')$. Since the
fields considered in this paper are mainly beams which
 propagate close to the $z$-direction,  we find
 convenient to introduce a finite quantization length
$L$ along the $z$ axis with discrete wave vector Cartesian
$z$-components $k_z \rightarrow \zeta_n = \frac{2 \pi}{L} n, \,
(n=0,  1 ,\ldots)$. The choice $n \geq 0$ implies that only the
parts of the field which propagate in the positive $z$-direction
are included in Eq. (\ref{a20}). The integral with respect to
$\mathrm{d} k_z$ in Eq. (\ref{a20}) is then replaced by a sum
\cite{LeeBook,Blow90a} $ \int \mathrm{d} k_x \mathrm{d} k_y
\mathrm{d} k_z \rightarrow \frac{2 \pi}{L} \sum_n \int
\mathrm{d}^2 \bk_T $, where  $\bk_T = ( k_x,  k_y)$ and
$\mathrm{d}^2 \bk_T \equiv \mathrm{d} k_x \mathrm{d} k_y$.
Moreover, we scale the annihilation and creation operators by
defining $\ha_\lambda(\bk) = \sqrt{L/(2 \pi)}
\ha_\lambda(\bk_T,n)$, in such a way that
\begin{equation}\label{a100}
[ \ha_\lambda(\bk_T,n), \ha_{\lambda'}^\dagger(\bk'_T,n')] =
\delta_{\lambda \lambda'} \delta_{n n'} \delta^{(2)}(\bk_T -
\bk'_T),
\end{equation}
where the Kronecker symbol replaces the delta function according
to $ \delta(k_z - k_z') \rightarrow \frac{L}{2 \pi} \delta_{n
n'}$.

A paraxial field is usually expressed as an envelope field
modulating a carrier plane wave with wave-vector $\bk_0$ and
angular frequency $\omega_0 = c |\bk_0|$. Without lack of
generality we assume $\bk_0 = k_0 \hat{\bz} $ and we choose
$k_0>0$ so that the carrier plane wave propagates  in the positive
$z$ direction.
By using the trivial identity
\begin{equation}\label{a420}
1 = \frac{\mathcal{L}}{2 \pi}\int_0^{2 \pi /\mathcal{L}}
\mathrm{d} k_0 \frac{\exp(\im k_0 z - \im \omega_0 t)}{\exp(\im
k_0 z - \im \omega_0 t)},
\end{equation}
where $\mathcal{L}$ is an arbitrary length,  we can rewrite the
vector potential operator as
\begin{equation}\label{a430}
\begin{array}{rcl}
\displaystyle{ \abA^{(+)}(\br,t) } &  = & \displaystyle{
 \abA^{(+)}(\br,t)  \times 1}
\\
 &  = &
\displaystyle{ \frac{\mathcal{L}}{2 \pi}\int_0^{2 \pi
/\mathcal{L}} \mathrm{d} k_0 \exp(\im k_0 z - \im \omega_0 t)
\hat{\bm \Psi}(\br,t)},
\end{array}
\end{equation}
where we have introduced the \emph{envelope} field
\begin{equation}\label{a120}
\begin{array}{rcl}
  \displaystyle{ \enve (\br,t)  } & = & \displaystyle{\sum_n \int \mathrm{d}^2
  \bk_T
  \left( \frac{\hbar}{8 \pi^2 \varepsilon_0 c |\bk| L} \right)^{1/2}}\\
  & &
    \displaystyle{\times  \sum_{\lambda = 1,2} \vepsilon^{(\lambda)}(\bk)
  \hat{a}_\lambda(\bk_T,n) }\\
  & &
      \displaystyle{\times  \exp \left[ \im ( \bk - k_0 \hat{\bz}) \cdot \br - \im  ( c|\bk| - \omega_0) t
  \right] },
\end{array}
\end{equation}
where  $\bk = \bk_T + \hat{\bz} \zeta_n$. Now, the key idea is to
find a subspace of the three-dimensional wave vector $\bk$-space
where the initial time envelope field $\enve (\br) \equiv \enve
(\br,t=0)$
 satisfies the  paraxial wave equation \cite{MandelBook}
\begin{equation}\label{a130}
\frac{\partial^2 \enve(\br)}{\partial x^2} + \frac{\partial^2
\enve(\br)}{\partial y^2} + 2 \im k_0 \frac{\partial
\enve(\br)}{\partial z} = 0.
\end{equation}
 If we substitute from Eq. (\ref{a120}) into Eq. (\ref{a130}) we
obtain
\begin{equation}\label{a150}
{\zeta_n }/{k_0}= 1 - {|\bk_T|^2}/{2 k_0^2}.
\end{equation}
This generalized dispersion relation plays a key role through this
paper. It defines a two-dimensional domain in the $\bk$-space
where \emph{both} d'Alembert wave equation and paraxial wave
equation are satisfied. For sake of clarity we write $\bq =
\bk_T$, $q = |\bq|$ and define the dimensionless parameter
$\vartheta = {q}/{( \sqrt{2 k_0^2}}) $.
If we denote with $\theta \in [0, \pi/2]$ the angle between the
wave vector $\bk$ and the axis $z$, then  $q/ \bk \cdot \hat{\bz}
= \tan \theta$ and from Eq. (\ref{a150}), it follows that
$\vartheta\sqrt{2}
 = -\cot \theta + \sqrt{(2 + \cot^2 \theta)}$. This
relation is exact; however, we can gain some insight if we
consider it in the limit $\theta \ll 1$ where $\vartheta \sqrt{2 }
\simeq \theta - {\theta^3}/{6} +
 O(\theta^5)$. This equations shows that $\vartheta \sqrt{2} = q/k_0$ is approximatively equal to
the \emph{divergence angle} of a Gaussian beam \cite{SiegmanBook}.
Moreover, by comparing $\vartheta$  with Eq. (2.8) by Deutsch and
Garrison \cite{Deutsch91a}, one recognizes $ \vartheta \sqrt{2}\ll
1$ as their perturbative expansion parameter. However, in our case
the only constraint is $\vartheta \leq 1$, as follows from Eq.
(\ref{a150}) and the condition $\zeta_n \geq 0$.

From Eq. (\ref{a150}) it readily follows that the exponential
factor in Eq. (\ref{a120}) can be written
\begin{equation}\label{a170}
\begin{array}{l}
  \displaystyle{ \exp \left[ \im ( \bk - k_0 \hat{\bz}) \cdot \br - \im (  c|\bk| - \omega_0) t
  \right]  }\\
  \displaystyle{= \exp (\im \bq \cdot \bx - \im \vartheta^2 k_0 z) \exp [-\im \omega_0 t (\sqrt{1+\vartheta^4}-1)]},
\end{array}
\end{equation}
where $\bx \equiv (x,y)$.  Clearly Eq. (\ref{a150}) affects also
the value of  the polarization unit vectors in Eq. (\ref{a120}).
To see this, we first write the total wave vector $\bk$ in terms
of $\bq$ and $k_0$ as $  \bk = \hat{\bq}q  + \hat{\bz}k_0(1 -
\vartheta^2)$, where $\hat{\bq} \equiv \bq/ |\bq|$, then we
arbitrarily choose (we always have this freedom)
$\vepsilon^{(2)}(\bq,\vartheta) = \hat{\bz} \times  \hat{\bq}$.
The remaining unit vector $\vepsilon^{(1)}$ is then uniquely fixed
by the cyclic relation $\vepsilon^{(1)} \propto \vepsilon^{(2)}
\times \bk$ to the value
%
%\begin{equation}\label{a220}
%\vepsilon^{(1)}(\bq,\vartheta) = \frac{(1 - \vartheta^2)\hat{\bq} -
%\sqrt{2} \vartheta \hat{\bz}}{(1 + \vartheta^4)^{1/2}}.
%\end{equation}
%
%
\begin{equation}\label{a220}
\vepsilon^{(1)}(\bq,\vartheta) = [ \hat{\bq}(1 - \vartheta^2) -
 \hat{\bz} \vartheta \sqrt{2} ]/(1 + \vartheta^4)^{1/2}.
\end{equation}
It is now possible to write explicitly the envelope field $\enve
(\br)$ restricted to the  $\bk$-subspace defined by the dispersion
relation Eq. (\ref{a150}). From the definition of $\zeta_n$ and
Eq. (\ref{a150}), it follows that we must select from the sum over
$n$ in Eq. (\ref{a120}) only those terms corresponding to $n =
n(\vartheta) \equiv \left[ \frac{k_0 L}{2 \pi}\left( 1 -
\vartheta^2 \right)\right]_\mathrm{IP}$, where ``IP'' stands for
Integer Part. This objective can be achieved by replacing in Eq.
(\ref{a120}) $\sum_n \rightarrow \sum_n \delta_{n, n(\vartheta)}$,
where $\delta_{n, n(\vartheta)} =1$ for $n = n(\vartheta)$ and
$\delta_{n, n(\vartheta)} =0$ otherwise. Finally, we can write
from Eq. (\ref{a120})
\begin{equation}\label{a280}
\begin{array}{rcl}
  \displaystyle{ \enve (\br)  } & = &\displaystyle{ \sum_n \int \mathrm{d}^2
  \bq
  \left( \frac{\hbar}{8 \pi^2 \varepsilon_0 \omega_0(1 + \vartheta^4)^{1/2} L} \right)^{1/2}}\\
 & &
    \displaystyle{\times  \sum_{\lambda = 1,2} \vepsilon^{(\lambda)}(\bq, \vartheta)
  \hat{a}_\lambda(\bq,n)\delta_{n, n(\vartheta)} }\\
  & &
      \displaystyle{\times   \exp (\im \bq \cdot \bx - \im \vartheta^2 k_0 z) },
\end{array}
\end{equation}
where $\vepsilon^{(\lambda)}(\bq,\vartheta)$ are given by Eq.
(\ref{a220}) and previous formulae. We have substituted everyplace
in  Eq. (\ref{a280})  $\zeta_n$ by Eq. (\ref{a150}); this
operation is permitted by the presence of $\delta_{n,
n(\vartheta)}$ within the sum. The only exception to these
substitutions is represented by the operators
$\hat{a}_\lambda(\bq,n)$ which, for the moment, are left
unchanged.

At this point, we  note that since the restriction to the paraxial
$\bk$-subspace has already been achieved via Eq. (\ref{a280}), it
is possible to
 make a step backward from the discrete momentum
$\zeta_n$ to the  continuous frequency  $\omega$:
%
%\begin{equation}\label{a405}
%c \zeta_n = \frac{2 \pi c}{L} n \rightarrow  \omega.
%\end{equation}
%
$c \zeta_n = \frac{2 \pi c}{L} n \rightarrow  \omega$. The other
required replacements are $\sum_n \rightarrow \frac{L}{2 \pi c}
\int \mathrm{d} \omega$, $\hat{a}_\lambda(\bq,n) =\sqrt{\frac{2
\pi c }{L}}\hat{a}_\lambda(\bq,\omega)$ and
\begin{equation}\label{a405}
\delta_{n, n(\vartheta)} = \frac{2 \pi c}{L} \delta\left[\omega -c
k_0 \left(1 - \vartheta^2\right) \right],
\end{equation}
 in such a way that
\begin{equation}\label{a410}
[\ha_\lambda(\bq, \omega), \ha_{\lambda'}^\dagger(\bq', \omega')]
=\delta_{\lambda \lambda'} \delta(\omega - \omega')
\delta^{(2)}(\bq - \bq').
\end{equation}
 Equation (\ref{a430}) can be then written as
\begin{widetext}
\begin{equation}\label{a520}
\begin{array}{rcl}
  \displaystyle{
   \abA^{(+)}(\br,t) }& =& \displaystyle{ \left( \frac{\mathcal{L}}{L} \right) \int_0^{2 \pi / \mathcal{L}} \mathrm{d} k_0
   \exp(\im k_0 z - \im \omega_0 t)    \int \mathrm{d} \omega \int \mathrm{d}^2
  \bq  \left( \frac{\hbar}{16 \pi^3 \varepsilon_0 k_0 (1 + \vartheta^4)^{1/2}}
  \right)^{1/2}}\\
 & \times & \displaystyle{
 \sum_{\lambda = 1,2} \vepsilon^{(\lambda)}(\bq,\vartheta)
 \hat{a}_\lambda(\bq,\omega) \delta\left[\omega - c k_0 \left(1 - \vartheta^2\right) \right]\exp (\im \bq \cdot \bx - \im \vartheta^2 k_0 z)
\exp [-\im \omega_0 t (\sqrt{1+\vartheta^4}-1)] }\,\
\end{array}
\end{equation}
\end{widetext}
where the first term $\mathcal{L}/{L}$ is a dimensionless constant
factor which can be eliminated by renormalizing
$\abA^{(+)}(\br,t)$ in the end of the calculations; therefore we
leave it out from our formulas.
 This is  not yet our final
expression since we can perform explicitly the integration with
respect to $k_0$ by using the well known formula for a delta of a
function which gives us
%$\delta\left[\omega - c k_0 \left(1 -
%\vartheta^2\right) \right] \rightarrow \frac{\delta \left( k_0 -
%\Omega_0 /c \right)}{1 + \Theta^2}$,
%
%\begin{equation}\label{a530}
%\delta\left[\omega - c k_0 \left(1 - \vartheta^2\right) \right]
%\rightarrow \frac{\delta \left( k_0 - \Omega_0 /c \right)}{1 +
%\Theta^2},
%\end{equation}
%
%
\begin{equation}\label{a530}
\delta\left[\omega - c k_0 \left(1 - \vartheta^2\right) \right]
\rightarrow {\delta \left( c k_0 - \Omega_0  \right)}/({1 +
\Theta^2}),
\end{equation}
where
% $\Theta \equiv {  \sqrt{2 q^2 c^2}}/[{\omega +
%\sqrt{\omega^2 + 2 q^2 c^2 }}]$,
%
\begin{equation}\label{a532}
\Theta \equiv {  \sqrt{2 q^2 c^2}}/[{\omega + \sqrt{\omega^2 + 2
q^2 c^2 }}],
\end{equation}
 $\Omega_0 /c \equiv
q/(\Theta\sqrt{2} )$
and only the term corresponding to $k_0 \geq 0$ has been retained.
Finally, we can write
\begin{widetext}
\begin{equation}\label{a540}
\begin{array}{rcl}
  \displaystyle{
   \abA^{(+)}(\br,t) }& =&  \displaystyle{
    \int \mathrm{d} \omega \exp[-\im \omega (t-z/c)]
     \left( \frac{\hbar}{16 \pi^3 \varepsilon_0 c \omega }
  \right)^{1/2}\int \mathrm{d}^2
  \bq   \sum_{\lambda = 1,2} \bm{\mathcal{E}}^{(\lambda)}(\bq, \omega, z,t)
 \hat{a}_\lambda(\bq,\omega) \exp \left(\im \bq \cdot \bx - \im \frac{q^2 c }{2
\Omega_0} z \right) },\\
\end{array}
\end{equation}
\end{widetext}
where we have defined the slowly varying polarization vectors
\begin{equation}\label{a543}
\begin{array}{l}
\displaystyle{\bm{\mathcal{E}}^{(\lambda)}(\bq,\omega, z,t) }
\equiv  \displaystyle{ \vepsilon^{(\lambda)}(\bq,\Theta) \left(
\frac{\omega^2/ \Omega_0^2}{ (1 + \Theta^4)(1 + \Theta^2)^4}
\right)^{1/4}} \\\\ \; \; \; \; \; \; \; \; \; \;\;\times
\displaystyle{ \exp \left[ \im \left(\omega - \Omega_0 \sqrt{1 +
\Theta^4} \right)t - \im \left(\omega - \Omega_0 \right)z/c
\right]} ,
\end{array}
\end{equation}
and the unit vectors $\vepsilon^{(\lambda)}(\bq,\Theta)$ are given
by Eq. (\ref{a220}) and previous formulae with $\vartheta
\rightarrow \Theta$.
%
%\end{widetext}
%

Equation (\ref{a540}) is  the first main result of this work. It
represents a vector potential field operator  which is a bona fide
transverse field obeying the d'Alembert wave equation for any time
$t >0$, whose corresponding envelope field satisfies the paraxial
wave equation (\ref{a130}) at $t=0$. We stress that this
expression is \emph{exact}, no approximations were made.

Now, we are ready to address the problem of building the quantum
mechanical state describing  a photon in a paraxial beam.  To this
end, we first note that until now the creation operators $
\hat{a}_\lambda^\dagger (\bq,\omega) $ have passed untouched
through all our operations. In fact, they still satisfy the
canonical commutation relations Eq. (\ref{a410}). However, their
form is not the most suitable one to deal with paraxial fields;
therefore we introduce the Fourier-transformed operators
\cite{Blow90a,Abouraddy02a}
\begin{equation}\label{a600}
\hat{a}_\lambda (\bx,\omega) = \frac{1}{2\pi} \int \mathrm{d}^2
\bq \, \hat{a}_\lambda(\bq,\omega) \exp( \im \bq \cdot \bx),
\end{equation}
such that
% $ [\ha_\lambda(\bx, \omega),
%\ha_{\lambda'}^\dagger(\bx', \omega')] =\delta_{\lambda \lambda'}
%\delta(\omega - \omega') \delta^{(2)}(\bx - \bx')$.
%
\begin{equation}\label{a605}
[\ha_\lambda(\bx, \omega), \ha_{\lambda'}^\dagger(\bx', \omega')]
=\delta_{\lambda \lambda'} \delta(\omega - \omega')
\delta^{(2)}(\bx - \bx').
\end{equation}
 If we substitute Eq. (\ref{a600}) in
Eq. (\ref{a540}) we obtain, after some algebra,
\begin{equation}\label{a622}
\abA^{(+)}(\br,t) = \sum_{\lambda = 1}^2 \int \mathrm{d} \omega
\frac{ e^{ - \im \omega(t - z/c)} }{(4 \pi \varepsilon_0 c
\omega/\hbar)^{1/2} } { \; \; \, {\hat{\! \! \!\bm{  \mathcal{A}}
}^{(\lambda)}}} (\bx,z,\omega,t),
\end{equation}
where $\br = (\bx,z)$ and we have introduced the \emph{exact}
slowly varying photon annihilation vector operators
\begin{equation}\label{a606}
{ \; \; \, {\hat{\! \! \!\bm{  \mathcal{A}}
}^{(\lambda)}}}(\bx,z,\omega,t) \equiv \int \mathrm{d}^2
  \bx'
    \bm{\mathcal{F}}^{(\lambda)}(\bx, z, \bx', \omega , t)
    \hat{a}_\lambda(\bx',\omega).
\end{equation}
Moreover, in Eq. (\ref{a606}) we have defined the Maxwell-paraxial
(MP) slowly varying modes
\begin{equation}\label{a620}
\begin{array}{rcl}
  \displaystyle{
\bm{\mathcal{F}}^{(\lambda)}(\bx,z, \bx', \omega , t)}& =&
\displaystyle{\frac{1}{(2 \pi)^2}\int \mathrm{d}^2
  \bq \, \bm{\mathcal{E}}^{(\lambda)}(\bq,\omega, z,t)}\\
  & & \times
\displaystyle{ \exp \left[\im \bq \cdot (\bx- \bx') - \im
\frac{q^2 c }{2 \Omega_0} z \right] }.
\end{array}
\end{equation}
Equation (\ref{a620}) displays the second main result of this
paper. It describes  the field in the plane $z$ at time $t$ due to
a point source with frequency $\omega$ located at $\bx'$ in the
transverse plane $z=0$. As its shape clearly suggests,
$\bm{\mathcal{F}}^{(\lambda)}(\bx, z, \bx', \omega, t)$ is the
quantum analog of the classical Huygens-Fresnel diffracted field.
This may be seen more clearly  by writing Eq. (\ref{a620}) in the
narrow-beam limit which is achieved by restricting the transverse
momentum integral to the domain $\mathcal{C}_\omega = \{ \bq : q
\ll \omega /c \}$. It is easy to see that within this domain
$\Theta \simeq \frac{q c}{\omega \sqrt{2}} \ll 1$, $\Omega_0
\simeq \omega$ and $ \bm{\mathcal{E}}^{(\lambda)}(\bq, \omega,
z,t) \simeq \ve^{(\lambda)}(\bq) $, where $\ve^{(1)}(\bq) =
\hat{\bq} + O(\Theta)$ and $\ve^{(2)}(\bq) = \hat{\bz} \times
\hat{\bq}$, are the zeroth order polarization unit vectors. In
this limit Eq. (\ref{a620}) reduces to the well known paraxial
Green's function
 $\bm{\mathcal{P}}^{(\lambda)}(\bx, z, \bx',
\omega)$ \cite{MandelBook}
\begin{equation}\label{a621}
\bm{\mathcal{P}}^{(\lambda)}(\bx, z, \bx', \omega) =
\int_{\mathcal{C}_\omega} \frac{\mathrm{d}^2 \bq}{(2 \pi)^2}
\ve^{(\lambda)}(\bq)  e^{\im \left[ \bq \cdot (\bx- \bx') -
\frac{q^2 c }{2 \omega} z \right]},
\end{equation}
here generalized to vector fields.
Now, it is straightforward to show that at a fixed time $t$, in
each transverse plane $z$, the MP functions
$\bm{\mathcal{F}}^{(\lambda)}(\bx, z, \bx', \omega , t)$ are
quasi-orthogonal
\begin{widetext}
\begin{equation}\label{a630}
\int \mathrm{d}^2 \, \bx \bm{\mathcal{F}}^{(\lambda)}(\bx, z,
\bx',\omega , t)
  \cdot \bm{\mathcal{F}}^{(\mu)}(\bx, z, \bx'' ,\omega , t)= \delta_{\lambda \mu}
   \int \frac{ \mathrm{d}^2 \bq}{(2 \pi)^2} \,
 \left( \frac{\omega^2/
\Omega_0^2}{ (1 + \Theta^4)(1 + \Theta^2)^4} \right)^{1/2}
\exp[\im \bq \cdot (\bx' -
  \bx'')],
\end{equation}
\end{widetext}
that is,  the right side of this equation approaches
$\delta_{\lambda \mu} \delta^{(2)}(\bx' - \bx'')$ in the
narrow-beam limit $\omega/(qc) \rightarrow \infty $, as expected
from the orthogonality of the classical paraxial Green's functions
for free space propagation. This result was already found by
Visser and Nienhuis \cite{Visser05a} who suggested to interpret
the Fourier-transformed creation operator
$\hat{a}_\lambda^\dagger(\bx, \omega)$  as the operator which
creates at $t = 0$ a photon with polarization $\lambda$ in the
paraxial mode $\exp(\im \omega z/c)
\bm{\mathcal{P}}^{(\lambda)}(\bx, z, \bx',\omega)$. More
generally, the interpretation of
$\bm{\mathcal{F}}^{(\lambda)}(\bx, z, \bx', \omega , t)$ as
single-photon  wave function can be put on a rigorous basis by
introducing the ``transverse-position'' states $| \bx,\omega ,
\lambda \ra \equiv \hat{a}_\lambda^\dagger(\bx, \omega) | 0 \ra $.
Then it readily follows that $\la 0 | \abA^{(+)}(\bx',z,t) |\bx
,\omega, \lambda \ra \propto \bm{\mathcal{F}}^{(\lambda)}(\bx', z,
\bx, \omega,t) \exp[-\im \omega (t-z/c)]$.

Now, the definition of $| \bx,\omega , \lambda \ra$ makes possible
to associate to any single-photon state $| \psi \ra$, its
corresponding Maxwell-paraxial  wave function $\psi_\lambda(\bx,
\omega) \equiv \la \bx ,\omega, \lambda | \psi \ra$. Then, for
example, the MP wave function associated to the plane wave state
$| \bq,\omega , \lambda \ra \equiv \hat{a}_\lambda^\dagger(\bq,
\omega) | 0 \ra $, is simply given by the Fourier relation $ \la
\bx,\omega' , \lambda' | \bq,\omega , \lambda \ra =\frac{\exp(\im
\bq \cdot \bx)}{2 \pi} \delta_{\lambda \lambda'} \delta(\omega -
\omega')$. More generally, for a given complete set of orthogonal
transverse functions $\psi_{n m }(\bx, \omega)$ as, e.g., the
Hermite- or the Laguerre-Gaussian beams \cite{SiegmanBook}, it is
possible to build the corresponding MP single-photon state as
\begin{equation}\label{a640}
| n,m , \omega, \lambda \ra = \int \mathrm{d}^2 \bx \, \psi_{n m
}(\bx, \omega) | \bx,\omega , \lambda \ra,
\end{equation}
where $ \la \bx,\omega' , \lambda' |  n,m , \omega, \lambda \ra =
\psi_{n m }(\bx, \omega) \delta_{\lambda \lambda'} \delta(\omega -
\omega')$.

This equation is our third and final main result: it represents
the \emph{exact} quantum-mechanical state of each of the photons
in the classical paraxial beam $\psi_{n m }(\bx, \omega)$.

In conclusion, a non-perturbative quantization scheme for
electromagnetic paraxial fields has been introduced. It relies on
the fact that it is possible to select  some initial field
configurations which are exact solutions of the paraxial wave
equation. These configurations are then evolved at later times
with the d'Alembert wave equation. In this way we were able to
find explicit and manageable expressions for the exact field in
both momentum [Eq. (\ref{a540})] and in position [Eqs.
(\ref{a622}-\ref{a620})] representation. Moreover we gave an
unambiguous definition for Maxwell-paraxial quantum states [Eq.
(\ref{a640})]. This quantization method also suggests that beyond
the paraxial case a whole class of other exact solutions of both
paraxial and d'Alembert wave equations could be found.
\begin{acknowledgments}
It is a pleasure to acknowledge Jorrit Visser and Gerard Nienhuis
for fruitful discussions. We also acknowledge support from the EU
under the IST-ATESIT contract. This project is also supported by
FOM.
\end{acknowledgments}

\end{document}